\documentclass[10pt]{blois07}
\usepackage{graphicx}
\usepackage{cite,./mcite}

\begin{document}
\title{Exclusive vector meson electroproduction at HERA}
\author{Aharon Levy}
\institute{Tel Aviv University and DESY}
\maketitle

\begin{abstract}
The latest results on exclusive vector meson electroproduction from
HERA are reviewed. In particular, the new high-statistics measurements
of the $\rho^0$ electroproduction are presented and compared to
several models. 
\end{abstract}

\section{Introduction}
\label{sec:intro}

\begin{figure}[h]
\begin{minipage}{0.5\columnwidth}
\hspace{-1cm}
\centerline{\includegraphics[width=0.7\columnwidth]{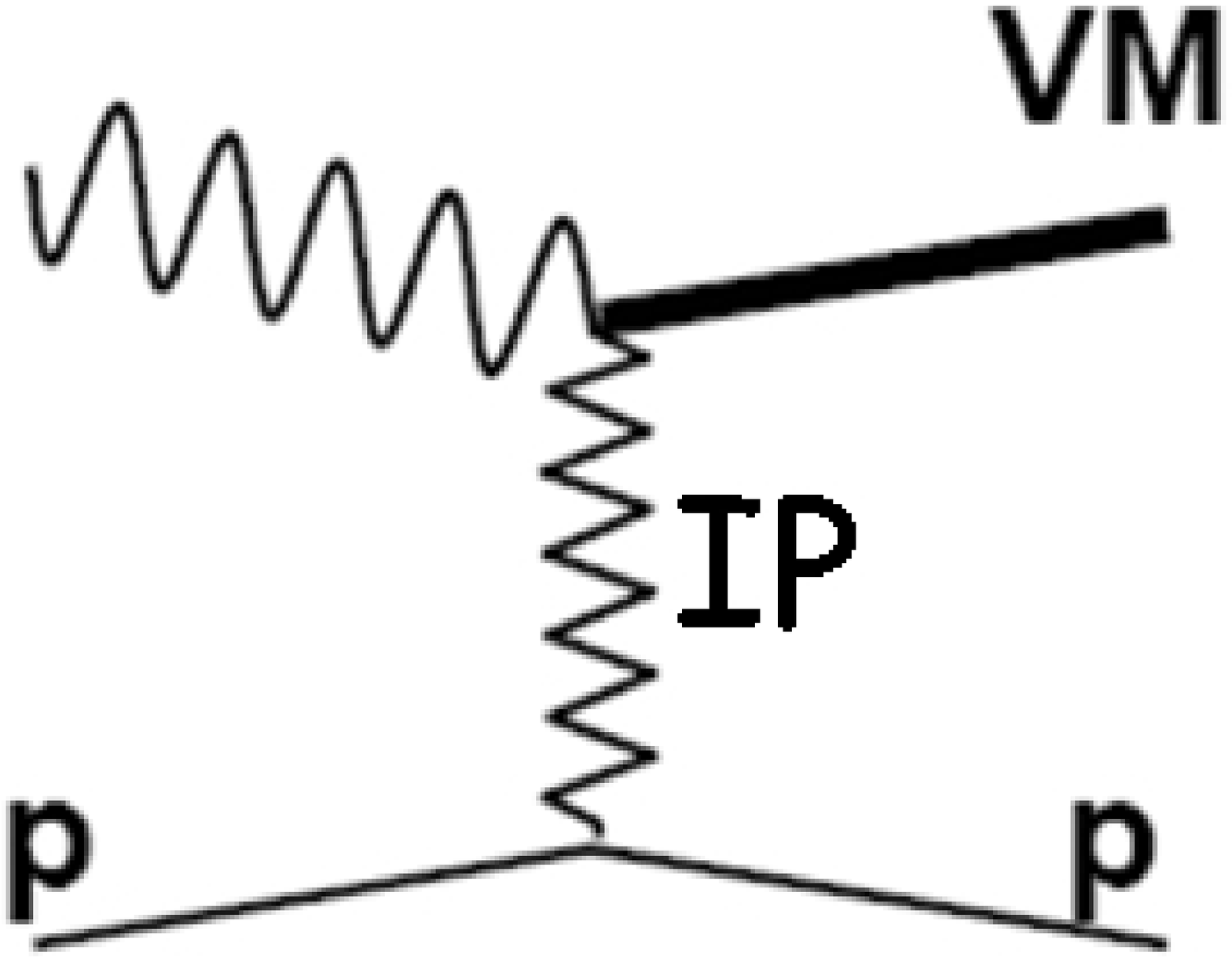}}
\caption{\it 
A diagram describing exclusive vector meson electroproduction in terms
of a Pomeron exchange.}
\label{fig:soft}
\end{minipage}
\hspace{2mm}
\begin{minipage}[h]{0.5\columnwidth}
\hspace{-0.2cm}
\centerline{\includegraphics[width=\columnwidth]{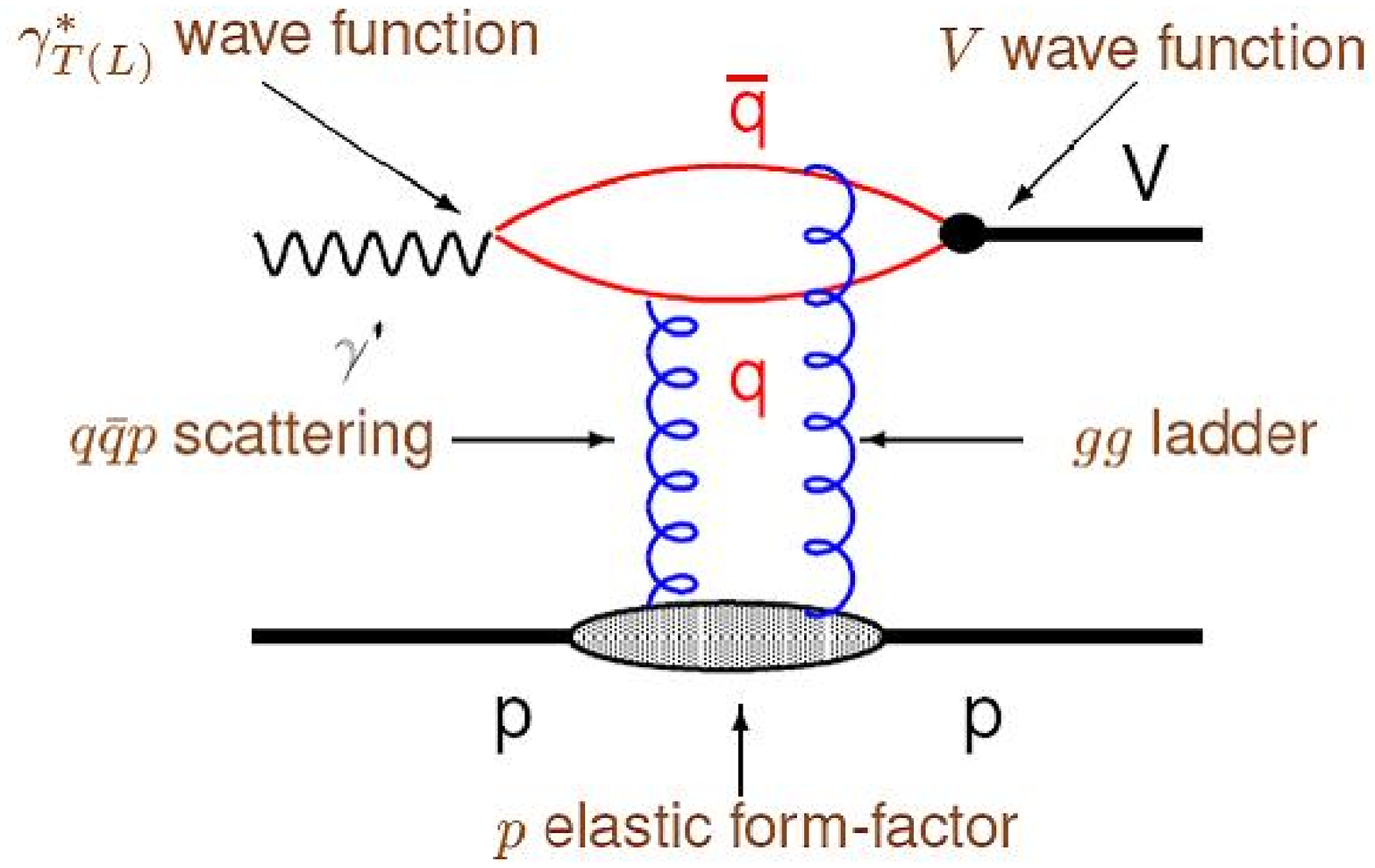}}
\caption{\it 
A diagram describing exclusive vector meson electroproduction in terms
of a two-gluon exchange.}
\label{fig:hard}
\end{minipage}
\end{figure}

Exclusive electroproduction of light vector mesons is a particularly
good process for studying the transition from the soft to the hard
regime of strong interactions, the former being well described within
the Regge phenomenology while the latter - by perturbative QCD
(pQCD). Among the most striking expectations~\cite{afs} in this
transition is the change of the logarithmic derivative $\delta$ of the
cross section $\sigma$ with respect to the $\gamma^* p$ center-of-mass
energy $W$, from a value of about 0.2 in the soft regime (represented
by a soft Pomeron~\cite{dl} exchange diagram in Fig.~\ref{fig:soft})
to 0.8 in the hard one (represented by a two-gluon exchange diagram in
Fig.~\ref{fig:hard}), and the decrease of the exponential slope $b$ of
the differential cross section with respect to the
squared-four-momentum transfer $t$, from a value of about 10
GeV$^{-2}$ to an asymptotic value of about 5 GeV$^{-2}$ when the
virtuality $Q^2$ of the photon increases.

When calculating the cross section of exclusive electroproduction of
vector mesons (V), one needs information on the wave function of the
initial virtual photon, the wave function of the produced vector
meson, the $q\bar{q}p$ scattering amplitude, which requires the gluon
density and the $p$ elastic form factor (see Fig.~\ref{fig:hard}).
In the following we present the available data on exclusive vector
meson electroproduction and then use the recent precision measurements
by ZEUS~\cite{zrho} of the $\rho^0$ vector meson to discuss what one
can learn about the produced vector meson wave function and about the
gluon density in the proton.

\section{$W$ dependence of the cross section}

\begin{figure}[h]
\hspace{-0.5cm}
\centerline{\includegraphics[width=0.8\columnwidth]{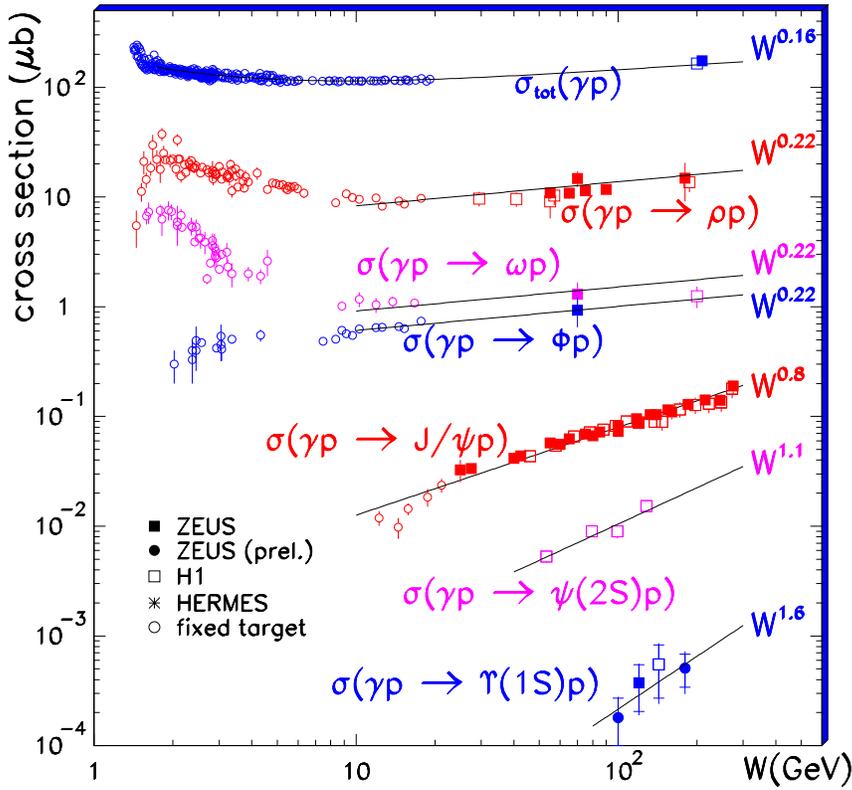}}
\vspace{-0.5cm}
\caption{\it 
The $W$ dependence of the cross section for exclusive vector meson
photoproduction, $\sigma(\gamma p \to V p)$. The total photoproduction
cross section is also shown. The lines are the result of a fit of the
form $ W^\delta$ to the high energy part of the data.}
\label{fig:sigvm}
\end{figure}

The soft to hard transition can be seen by studying the $W$ dependence
of the cross section for exclusive vector meson photoproduction, from
the lightest one, $\rho^0$, to the heavier ones, up to the
$\Upsilon$. The scale in this case is the mass of the vector meson, as
in photoproduction $Q^2$ = 0. Figure~\ref{fig:sigvm} shows
$\sigma(\gamma p \to V p)$ as function of $W$ for light and heavy
vector mesons. For comparison, the total photoproduction cross
section, $\sigma_{tot}(\gamma p)$, is also shown. The data at high $W$
can be parameterised as $W^\delta$, and the value of $\delta$ is
displayed in the figure for each reaction. One sees clearly the
transition from a shallow $W$ dependence for low scales to a steeper
one as the scale increases.

\begin{figure}[h]
\begin{minipage}{0.5\columnwidth}
\hspace{-0.5cm}
\centerline{\includegraphics[width=\columnwidth]{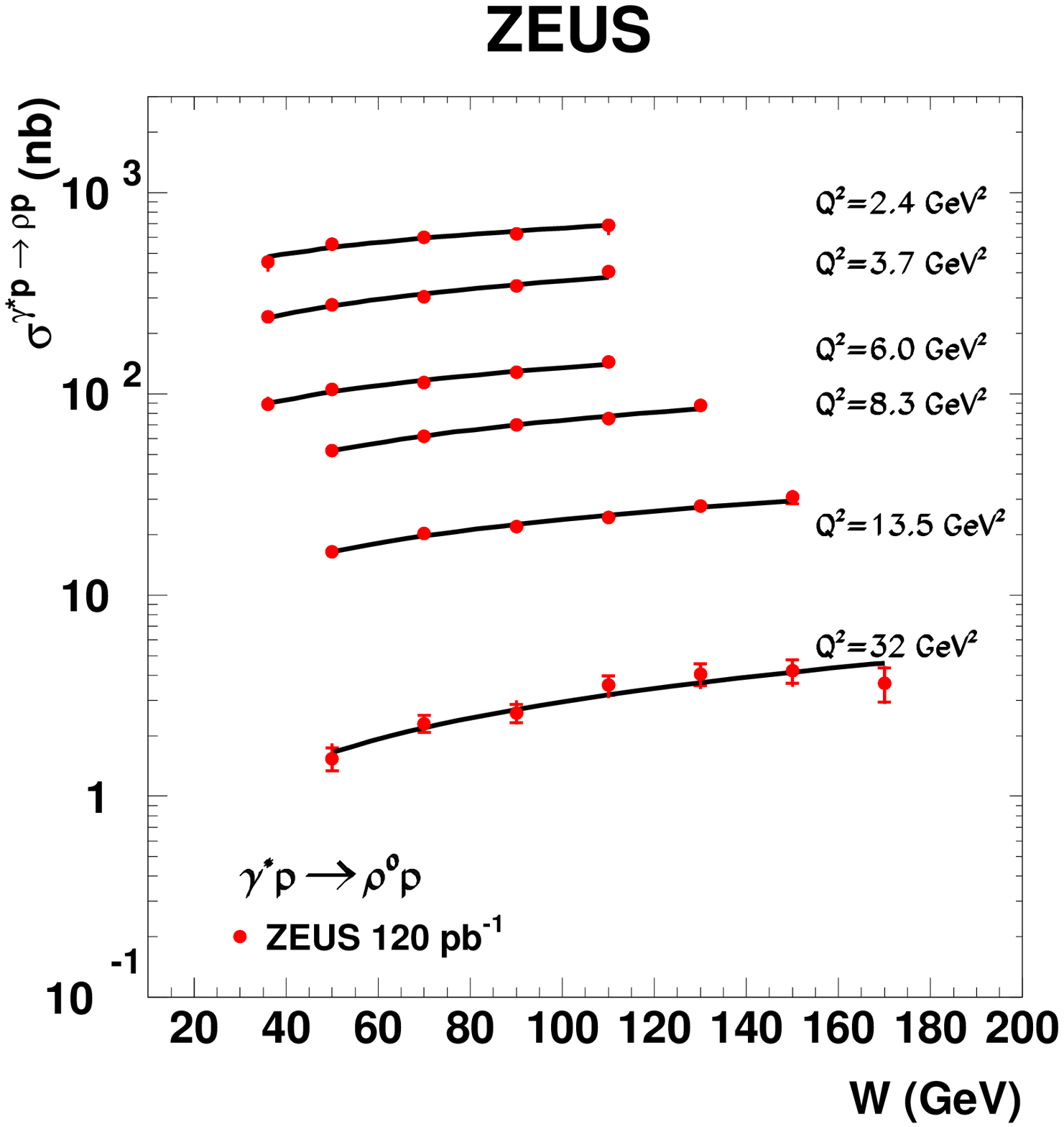}}
\vspace{-0.5cm}
\caption{\it 
The $W$ dependence of the cross section for exclusive $\rho^0$
electroproduction, for different $Q^2$ values, as indicated in the
figure.  The lines are the result of a fit of the form $ W^\delta$
to the data.}
\label{fig:w}
\end{minipage}
\hspace{2mm}
\begin{minipage}{0.5\columnwidth}
\hspace{-0.5cm}
\centerline{\includegraphics[width=1.25\columnwidth]{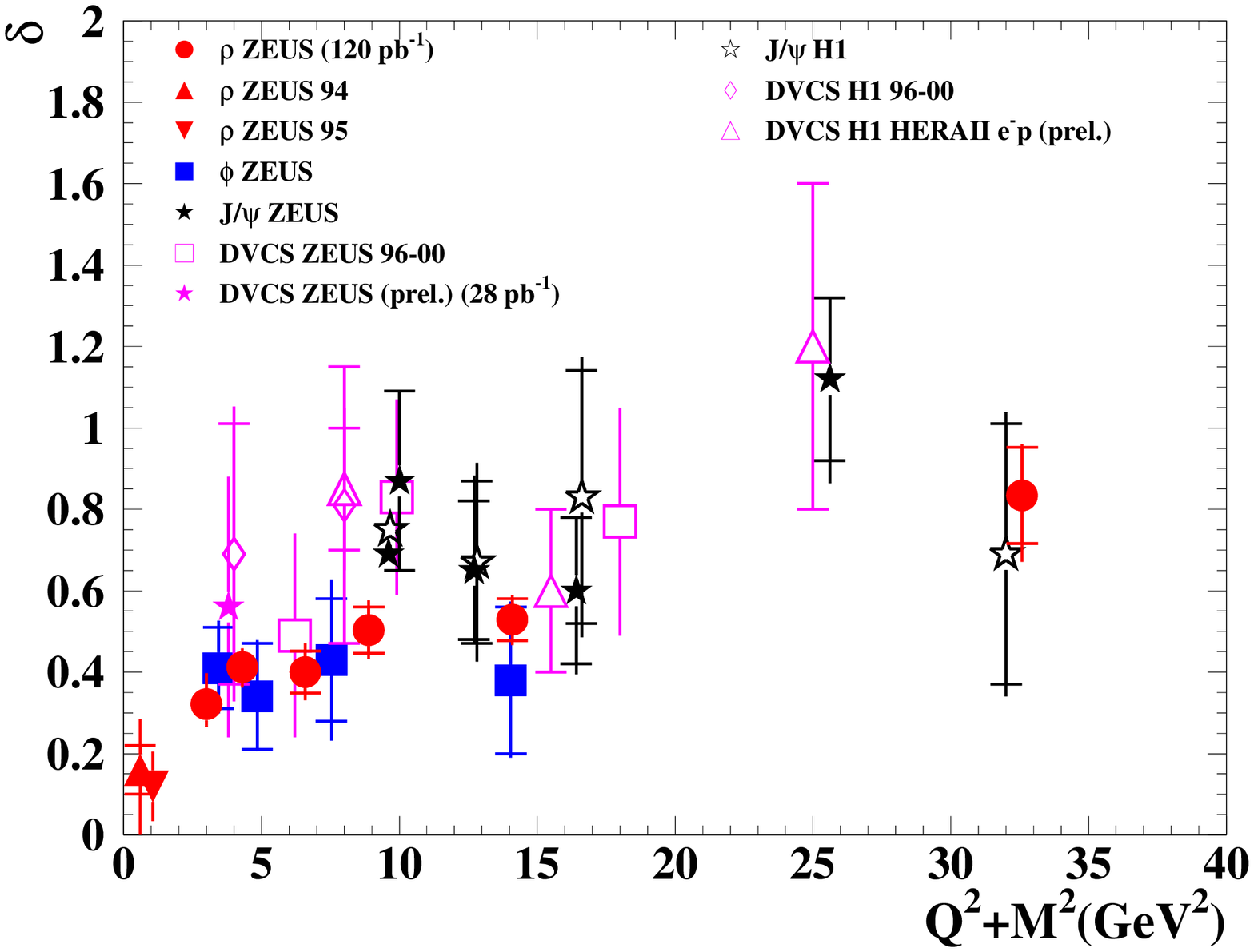}}
\vspace{-0.5cm}
\caption{\it 
A compilation of the value of $\delta$ from a fit of the form
$W^\delta$ for exclusive vector-meson electroproduction, as a function
of $Q^2+M^2$. It includes also the DVCS results.}
\label{fig:del07}
\end{minipage}
\end{figure}

One can also check this transition by varying $Q^2$ for a given vector
meson.  The cross section $\sigma (\gamma^* p \to \rho^0 p)$ is
presented in Fig.~\ref{fig:w} as a function of $W$, for different
values of $Q^2$. The cross section rises with $W$ in all $Q^2$ bins.
In order to quantify this rise, the logarithmic derivative $\delta$ of
$\sigma$ with respect to $W$ is obtained by fitting the data to the
expression $\sigma \sim W^\delta$ in each of the $Q^2$ intervals.  The
resulting values of $\delta$ from the recent ZEUS measurement are
compiled in Fig~\ref{fig:del07}.  Also included in this figure are
values of $\delta$ from other measurements~\cite{rho-other} for the
$\rho^0$ as well as those for $\phi$~\cite{phi}, $J/\psi$~\cite{jpsi}
and $\gamma$~\cite{dvcs} (Deeply Virtual Compton Scattering
(DVCS)). In this case the results are plotted as a function of
$Q^2+M^2$, where $M$ is the mass of the vector meson.  One sees a
universal behaviour, showing an increase of $\delta$ as the scale
becomes larger, in agreement with the expectations mentioned in the
introduction.  The value of $\delta$ at low scale is the one expected
from the soft Pomeron intercept~\cite{dl}, while the one at large
scale is in accordance with twice the logarithmic derivative of the
gluon density with respect to $W$.

\section{$t$ dependence of the cross section}

\begin{figure}[h]
\begin{minipage}{0.5\columnwidth}
\hspace{-0.5cm}
\centerline{\includegraphics[width=1.2\columnwidth]{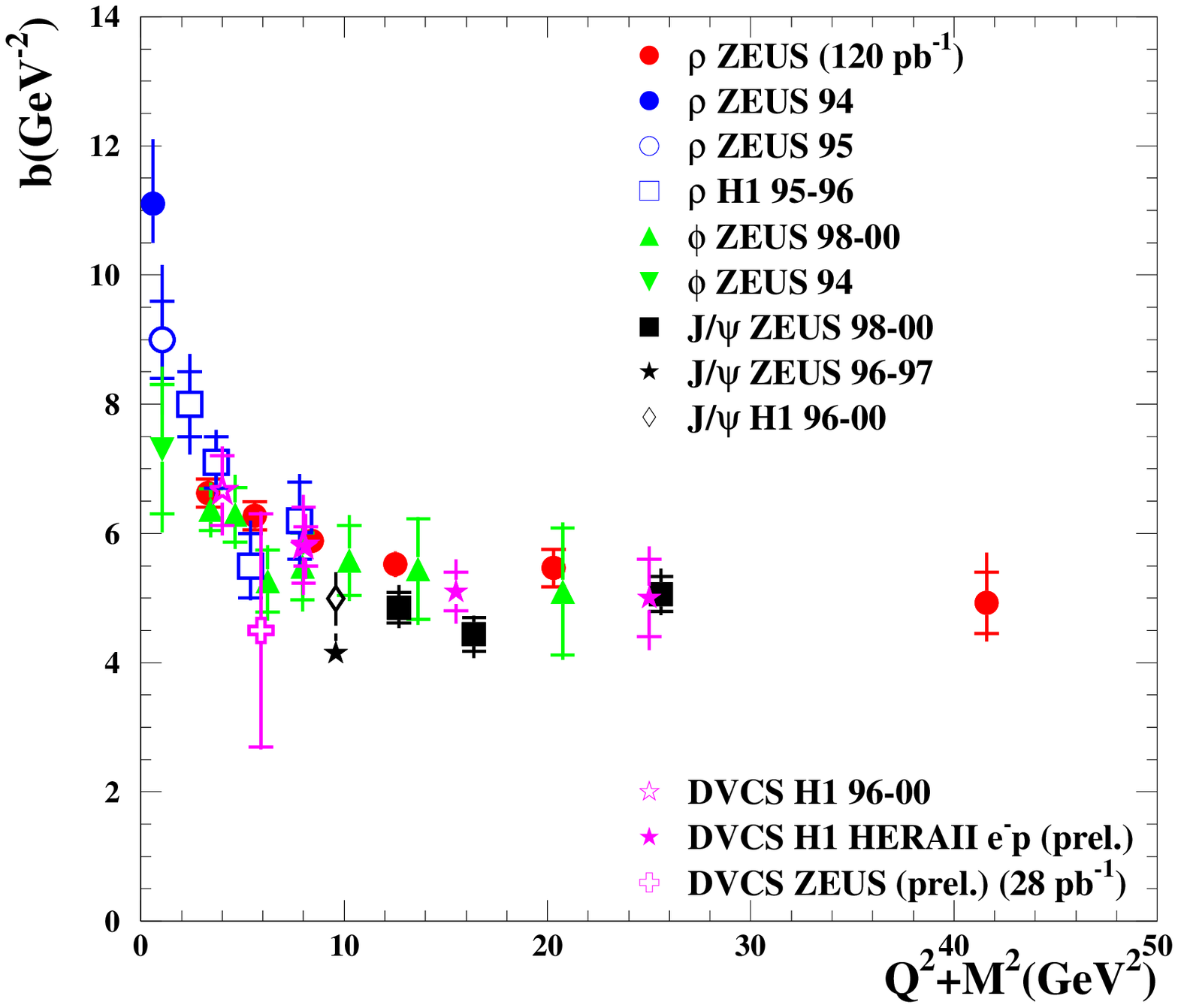}}
\vspace{-0.5cm}
\caption{\it 
A compilation of the value of the slope $b$ from a fit of the form
$d\sigma/d|t| \propto
e^{-b|t|}$ for exclusive vector-meson electroproduction, as a function
of $Q^2+M^2$. Also included is the DVCS result.}
\label{fig:b07}
\end{minipage}
\hspace{2mm}
\begin{minipage}{0.5\columnwidth}
\hspace{-0.5cm}
\centerline{\includegraphics[width=\columnwidth]{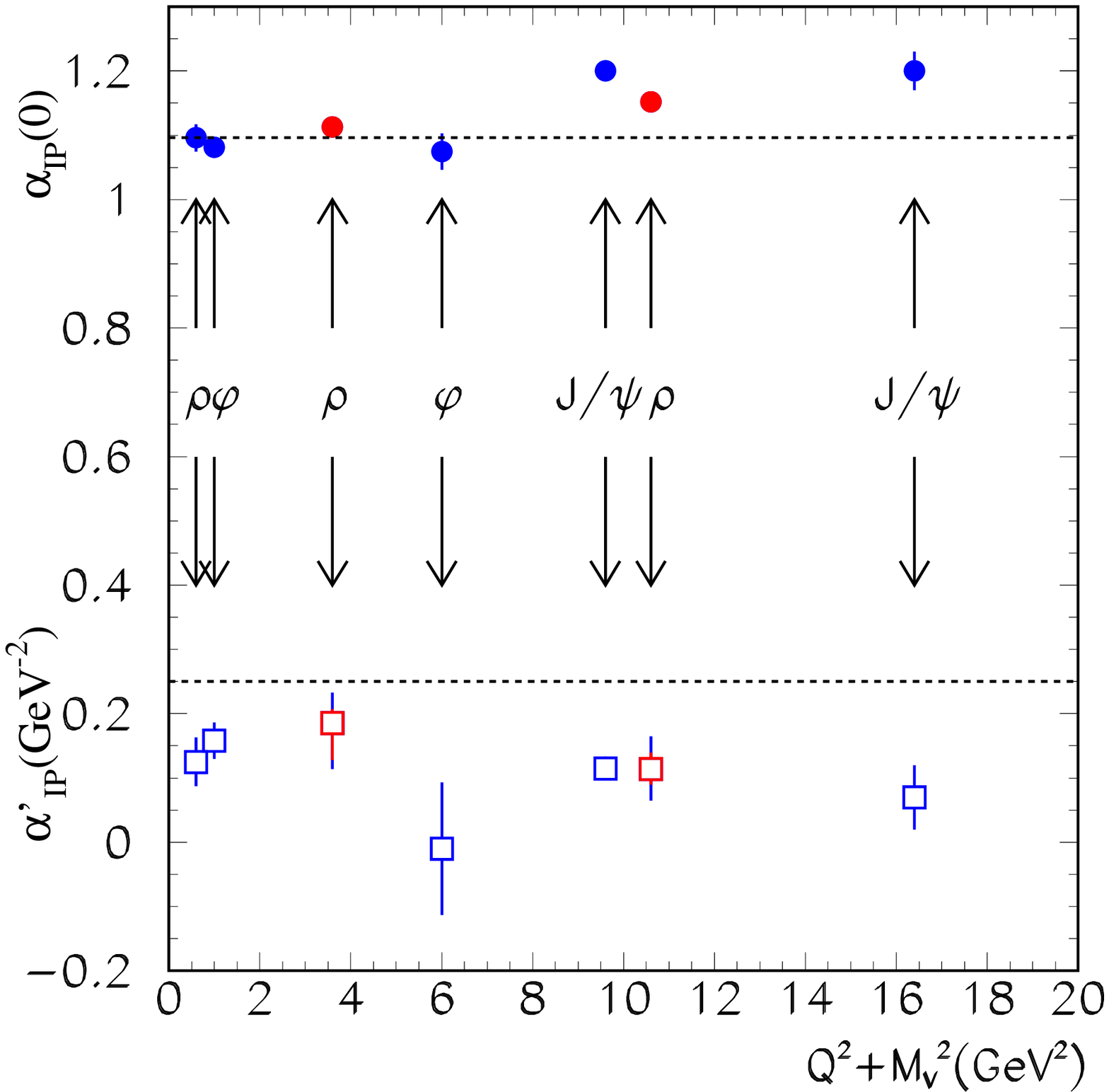}}
\vspace{-0.5cm}
\caption{\it 
Values of the intercept and slope of the effective Pomeron trajectory
as a function of $Q^2+M^2$, as obtained from measurements of exclusive
electroproduction of $\rho^0$, $\phi$, and $J/\psi$ vector mesons.}
\label{fig:ap-apr-pom}
\end{minipage}
\end{figure}

The differential cross section, d$\sigma$/d$t$, has been parameterised
by an exponential function $e^{-b|t|}$ and fitted to the data of
exclusive vector meson electroproduction and also to DVCS. The
resulting values of $b$ as a function of the scale $Q^2+M^2$ are
plotted in Fig.~\ref{fig:b07}. As expected, $b$ decreases to a
universal value of about 5 GeV$^{-2}$ as the scale increases.

The value of $b$ can be related via a Fourier transform to the impact
parameter. Assuming that the process of exclusive electroproduction of
vector mesons is hard and dominated by gluons, one can use the
relation $<r^2> = b (\hbar c)^2$ to obtain the radius of the gluon
density in the proton. The value of about 5 GeV$^{-2}$ corresponds to
a value of $<r>_g \sim$ 0.6 fm, smaller than the value of the charge density
of the proton ($ <r>_p \sim$ 0.8 fm), indicating that the gluons are
well-contained within the charge-radius of the proton.

One can study the $W$ dependence of d$\sigma$/d$t$ for fixed $t$
values and extract the effective Pomeron trajectory
$\alpha_{IP}(t)$. This was done in case of the $\rho^0$ for two $Q^2$
values and the trajectory was fitted to a linear form to obtain the
intercept $\alpha_{IP}(0)$ and the slope $\alpha_{IP}^\prime$. These
values are presented in a compilation of the effective Pomeron
intercept and slope, from this and from similar studies for other
vector mesons, in Fig.~\ref{fig:ap-apr-pom}. As in the other
compilations, the values are plotted as a function of $Q^2+M^2$.  The
value of $\alpha_{IP}(0)$ increases with $Q^2$ while the value of
$\alpha_{IP}^\prime$ tends to decrease with $Q^2$.

\section{$Q^2, W$ and $t$ dependence of $r_{00}^{04} = \sigma_L/\sigma_{tot}$ for 
$\gamma^* p \to \rho^0 p$}  

\begin{figure}[h]
\begin{minipage}{0.5\columnwidth}
\hspace{-0.5cm}
\centerline{\includegraphics[width=0.8\columnwidth]{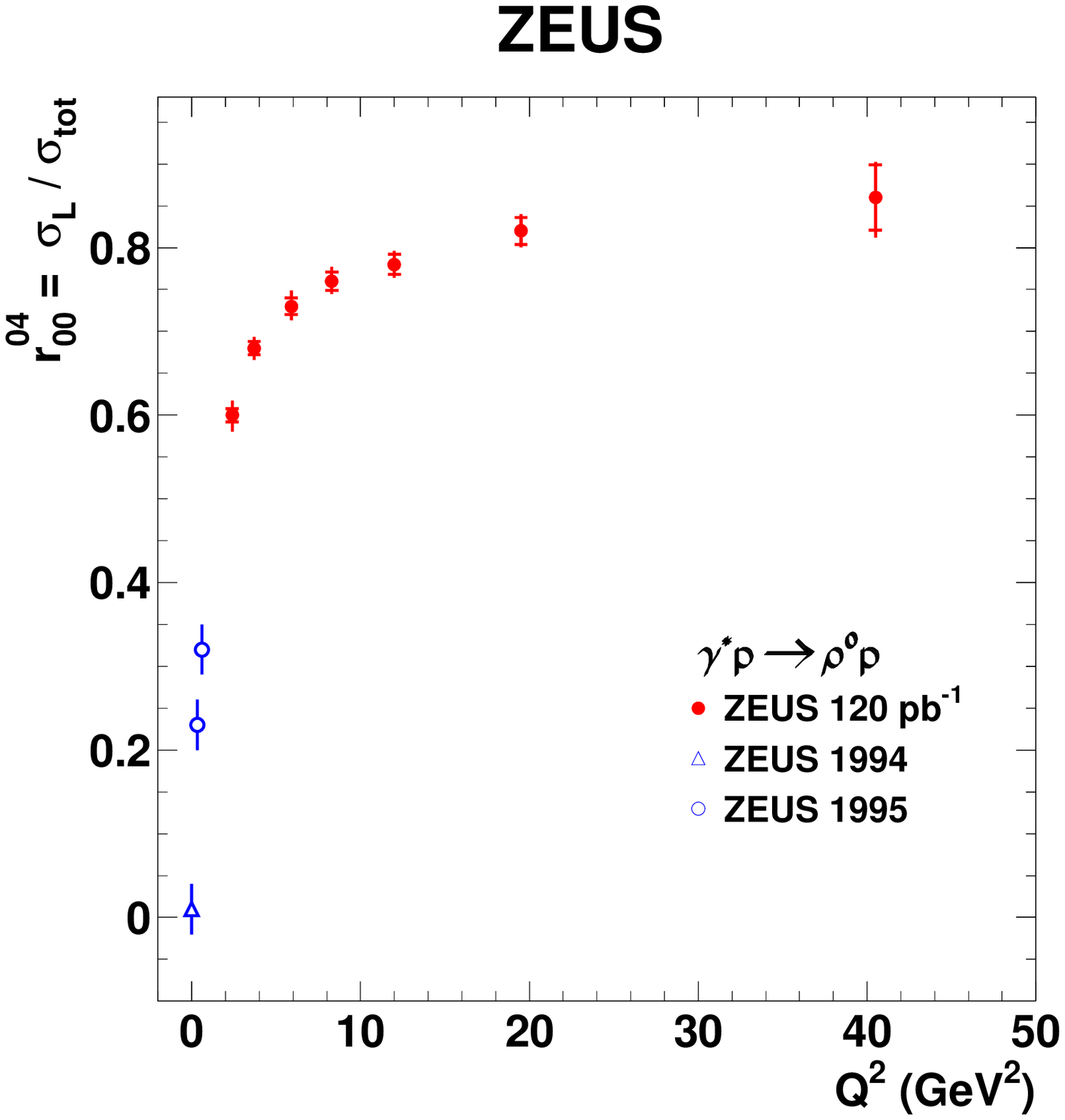}}
\vspace{-0.5cm}
\caption{\it 
The ratio $r^{04}_{00}$ as a function of $Q^2$ for $W =$ 90 GeV.}
\label{fig:r-q2}
\end{minipage}
\hspace{2mm}
\begin{minipage}{0.5\columnwidth}
\hspace{-0.5cm}
\centerline{\includegraphics[width=0.8\columnwidth]{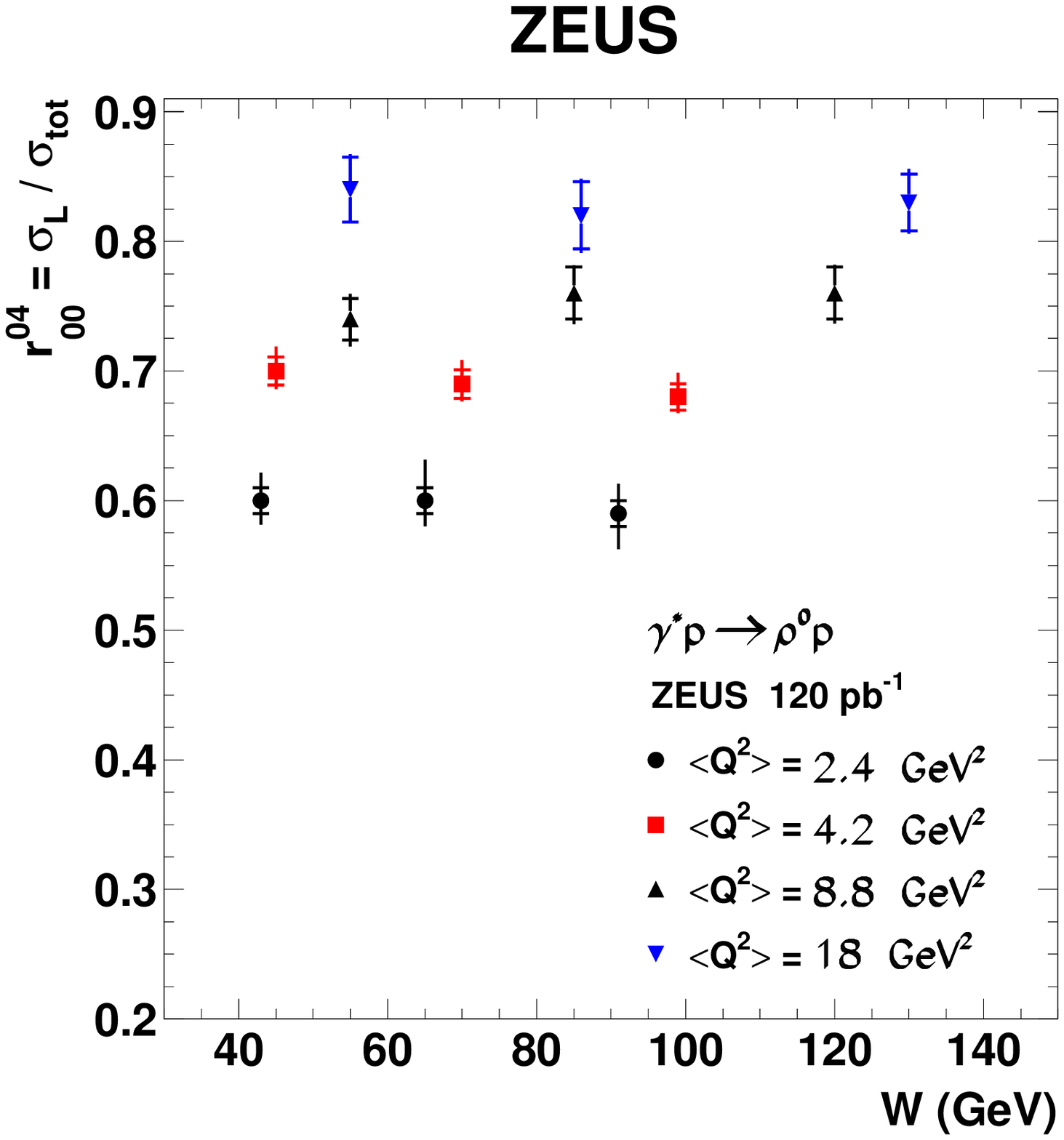}}
\vspace{-0.5cm}
\caption{\it 
The ratio $r^{04}_{00}$ as a function of $W$ for different values of $Q^2$, as
indicated in the figure.}
\label{fig:r-w}
\end{minipage}
\vspace{-0.5cm}
\begin{minipage}{0.5\columnwidth}
\hspace{-0.5cm}
\centerline{\includegraphics[width=0.8\columnwidth]{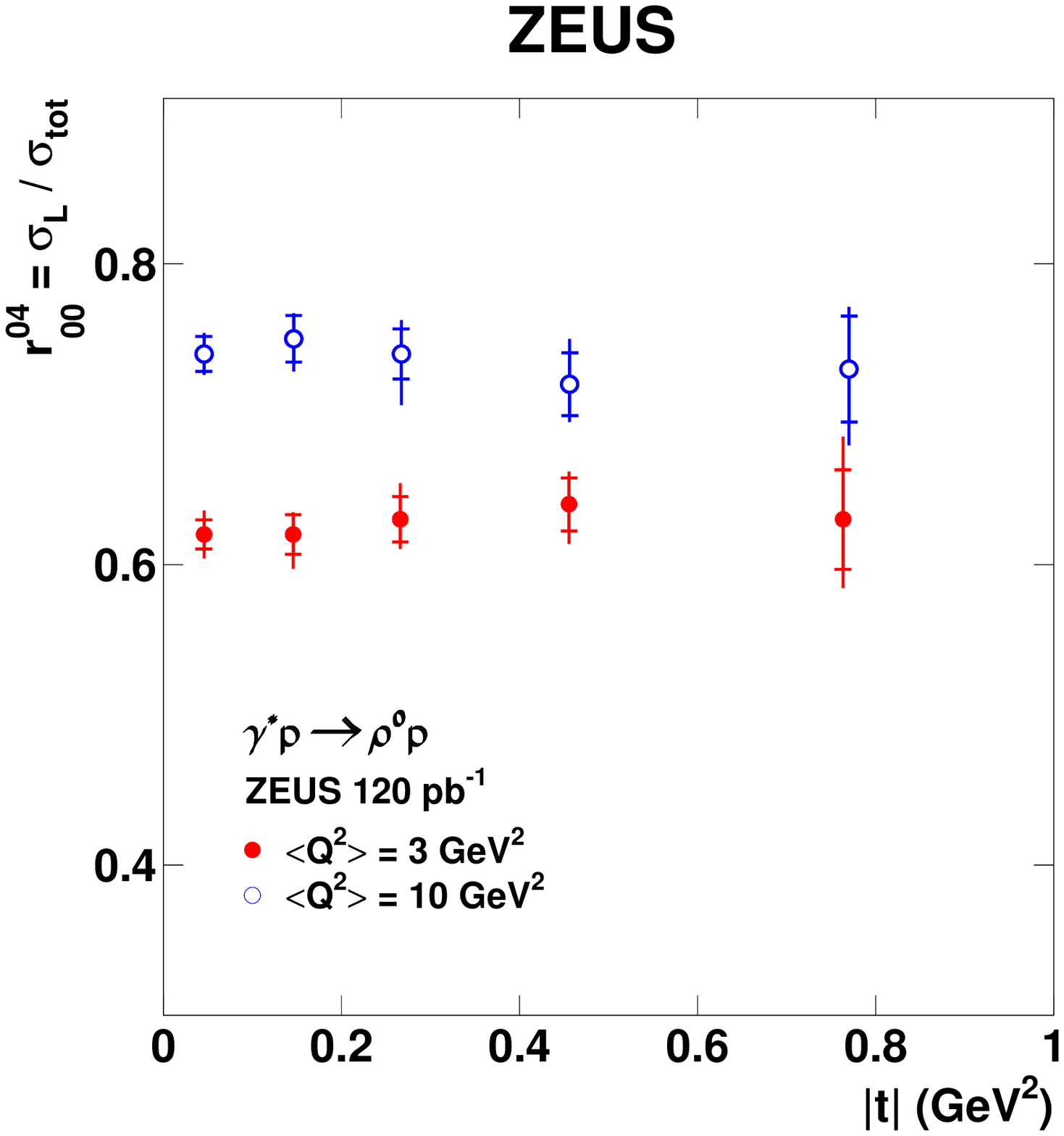}}
\vspace{-0.5cm}
\caption{\it 
The ratio $r^{04}_{00}$ as a function of $|t|$ for different values of $Q^2$, as
indicated in the figure.}
\label{fig:r-t}
\end{minipage}
\hspace{2mm}
\begin{minipage}{0.5\columnwidth}
\hspace{-0.5cm}
\centerline{\includegraphics[width=0.8\columnwidth]{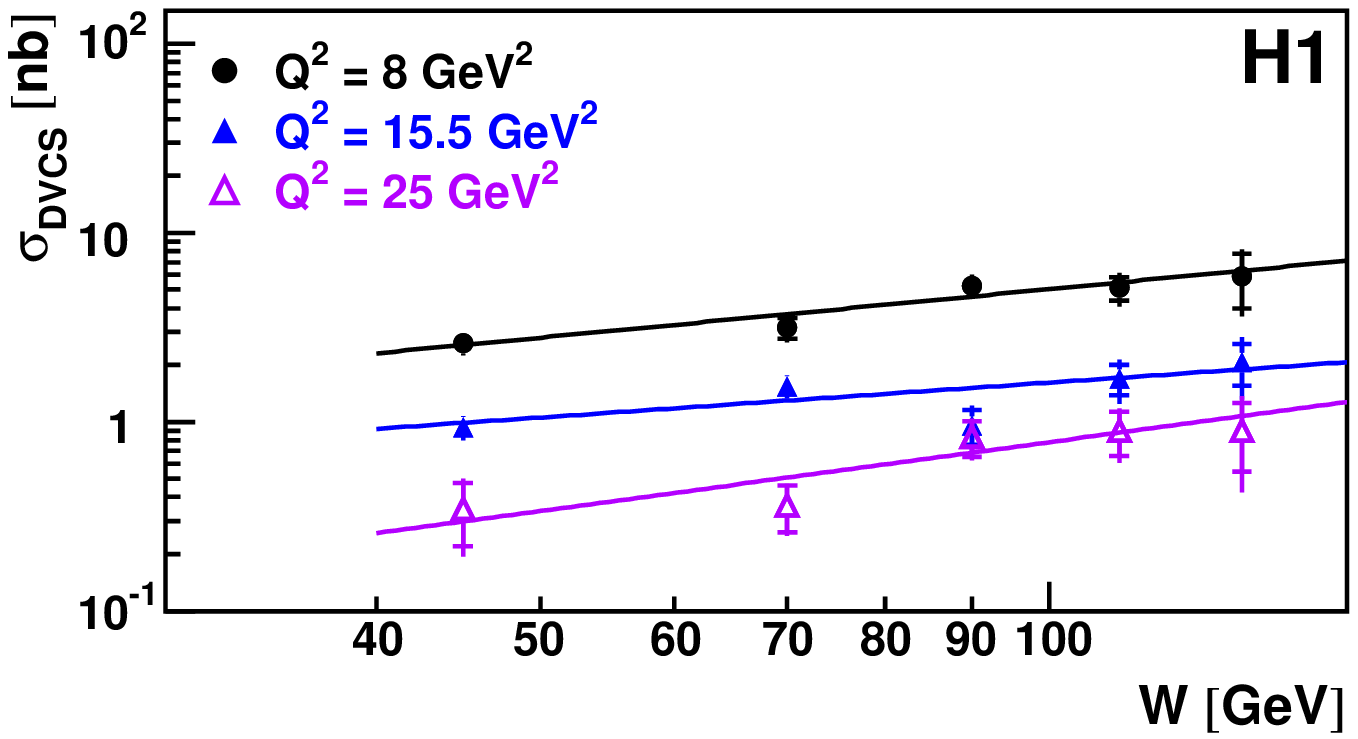}}
\vspace{-0.5cm}
\caption{\it 
The DVCS cross section as a function of $W$ at three values of
$Q^2$. The solid lines represent the results of fits of the form
$W^\delta$.}
\label{fig:dvcs}
\end{minipage}
\end{figure}

The helicity analysis of the decay-matrix element $r_{00}^{04}$ of
the $\rho^0$ was used to extract the ratio $R = \sigma_L/\sigma_T$ of
longitudinal ($\sigma_L$) to transverse ($\sigma_T$) $\gamma^* p$
cross sections. When the value of $r_{00}^{04}$ is close to one, as is
the case for this analysis, the error on $R$ becomes large and highly
asymmetric. It is then advantageous to study the properties of
$r_{00}^{04}$ itself which carries the same information ($=
\sigma_L/\sigma_{tot}$), rather than $R$. While $r_{00}^{04}$ is an increasing function 
of $Q^2$, as shown in Fig.~\ref{fig:r-q2}, it is independent of $W$ in
all $Q^2$ intervals (Fig.~\ref{fig:r-w}). This implies that the $W$
behaviour of $\sigma_L$ is the same as that of $\sigma_T$, a result
which is somewhat surprising. The $q\bar{q}$ configurations in the
wave function of $\gamma^*_L$ have typically a small transverse size,
while the configurations contributing to $\gamma^*_T$ may have large
transverse sizes.  The contribution to $\sigma_T$ of large-size
$q\bar{q}$ configurations, which are more hadron-like, is expected to
lead to a shallower $W$ dependence than in case of $\sigma_L$.  Thus,
the result presented in Fig.~\ref{fig:r-w} suggests that the
large-size configurations of the transversely polarised photon are
suppressed.

The above conclusion can also explain the behaviour of $r^{04}_{00}$
as a function of $t$, shown in Fig.~\ref{fig:r-t} for two $Q^2$
values. Different sizes of interacting objects imply different $t$
distributions, in particular a steeper $d\sigma_T/dt$ compared to
$d\sigma_L/dt$. This turns out not to be the case.  In both $Q^2$
ranges, $r^{04}_{00}$ is independent of $t$, reinforcing the earlier
conclusion about the suppression of the large-size configurations in
the transversely polarised photon.

This suppression is also seen in DVCS,
$\gamma^* p \to \gamma p$. The final state photon is real and
therefore transversely polarised. Using s-channel helicity
conservation, also the initial virtual photon would be transversely
polarised. Looking at the new H1 measurement of the DVCS cross
section~\cite{h1-dvcs}, shown in Fig.~\ref{fig:dvcs}, which has a
steep $W$ dependence ($\delta \sim$ 0.8), one concludes that the
large-size configurations in the transversely polarised photon are
suppressed.

\section{Comparison of the data of $\gamma^* p \to \rho^0 p$ to models}

The precision measurements of the reaction $\gamma^* p \to \rho^0 p$
were used to compare to some selected pQCD-inspired models.

All models are based on the dipole representation of the virtual
photon, in which the photon first fluctuates into a $q\bar{q}$ pair
(colour dipole), which then interacts with the proton to produce
the $\rho^0$. The ingredients necessary in such calculations are the
virtual-photon wave-function, the dipole-proton cross section and the
$\rho^0$ wave-function. The photon wave-function is known from QED.
The models differ in the treatment of the dipole-proton cross section
and the assumed $\rho^0$ wave-function.

The models of Frankfurt, Koepf and Strikman (FKS)~\cite{fks} and of
Martin, Ryskin and Teubner (MRT)~\cite{mrt} are based on two-gluon
exchange as the dominant mechanism for the dipole-proton interaction.
The gluon distributions are derived from inclusive measurements of the
proton structure function. In the FKS model, a three-dimensional
Gaussian is assumed for the $\rho^0$ wave-function, while MRT use
parton-hadron duality and normalise the calculations to the data. For
the comparison with the present measurements the MRST99~\cite{mrst99}
and CTEQ6.5M~\cite{cteq65m} parameterisations for the gluon density
were used.

Kowalski, Motyka and Watt (KMW)~\cite{kmw} use an improved version of
the saturation model~\cite{gbw}, with an explicit dependence on the
impact parameter and DGLAP evolution in $Q^2$, introduced through the
unintegrated gluon distribution~\cite{un-gl}. Forshaw, Sandapen and
Shaw (FSS)~\cite{fss} model the dipole-proton interaction through the
exchange of a soft~\cite{dl} and a hard~\cite{dl-hard} Pomeron, with
(Sat) and without (Nosat) saturation, and use the DGKP and Gaussian
$\rho^0$ wave-functions. In the model of Dosch and Ferreira
(DF)~\cite{df}, the dipole cross section is calculated using Wilson
loops, making use of the stochastic vacuum model for the
non-perturbative QCD contribution.

While the calculations based on two-gluon exchange are limited to
relatively high-$Q^2$ values (typically $\sim$ 4 GeV$^2$), those
based on modelling the dipole cross section incorporate both the
perturbative and non-perturbative aspects of $\rho^0$ production.

\begin{figure}[h]
\begin{minipage}{0.5\columnwidth}
\hspace{-0.5cm}
\centerline{\includegraphics[width=\columnwidth]{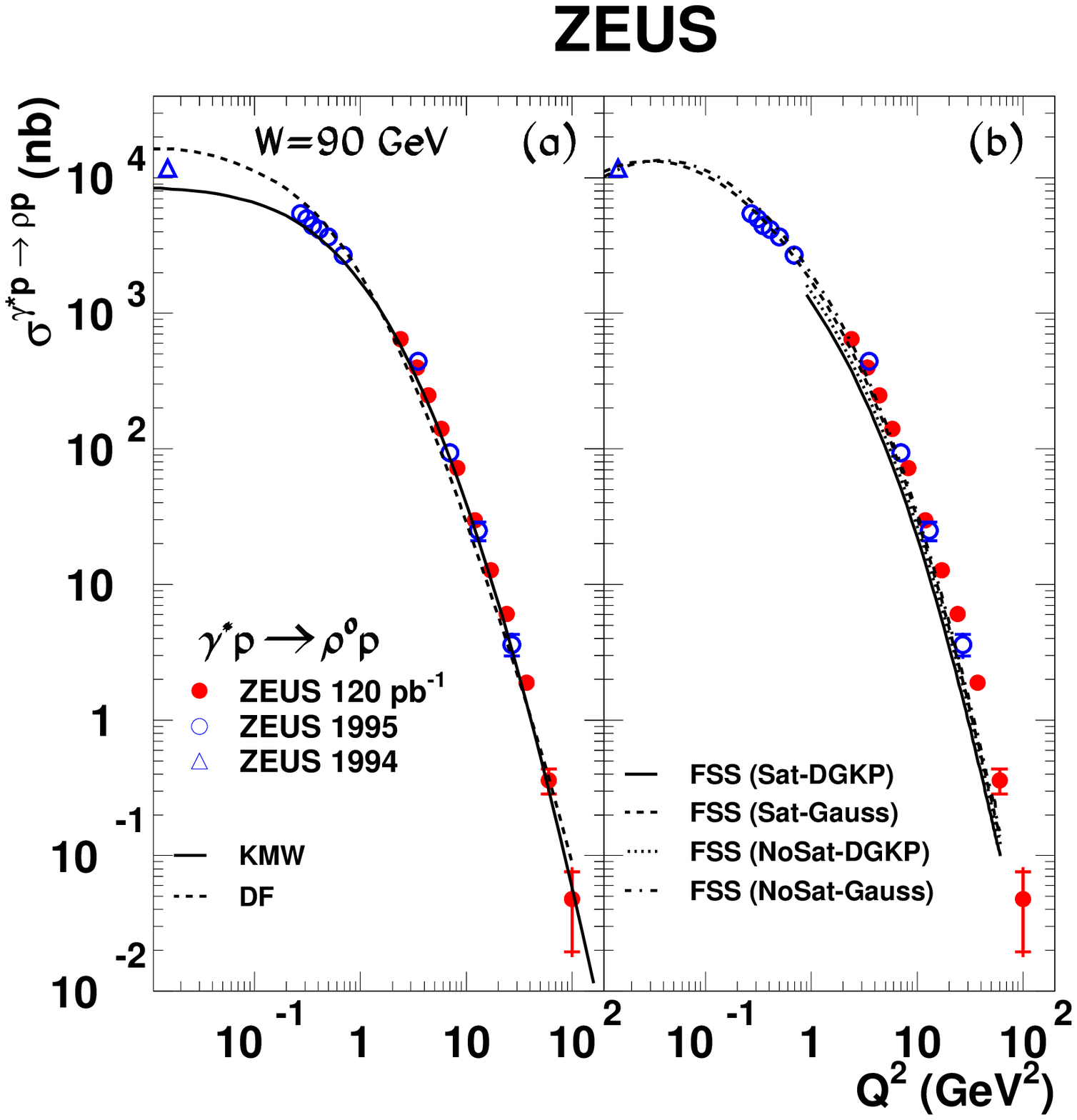}}
\vspace{-0.5cm}
\caption{\it
The $Q^2$ dependence of the $\gamma^* p \to \rho^0 p$ cross section at
$W$=90 GeV.  The same data are plotted in (a) and (b), compared to
different models, as described in the text. }
\label{fig:q2-mod}
\end{minipage}
\hspace{2mm}
\begin{minipage}[h]{0.5\columnwidth}
\hspace{-0.5cm}
\centerline{\includegraphics[width=\columnwidth]{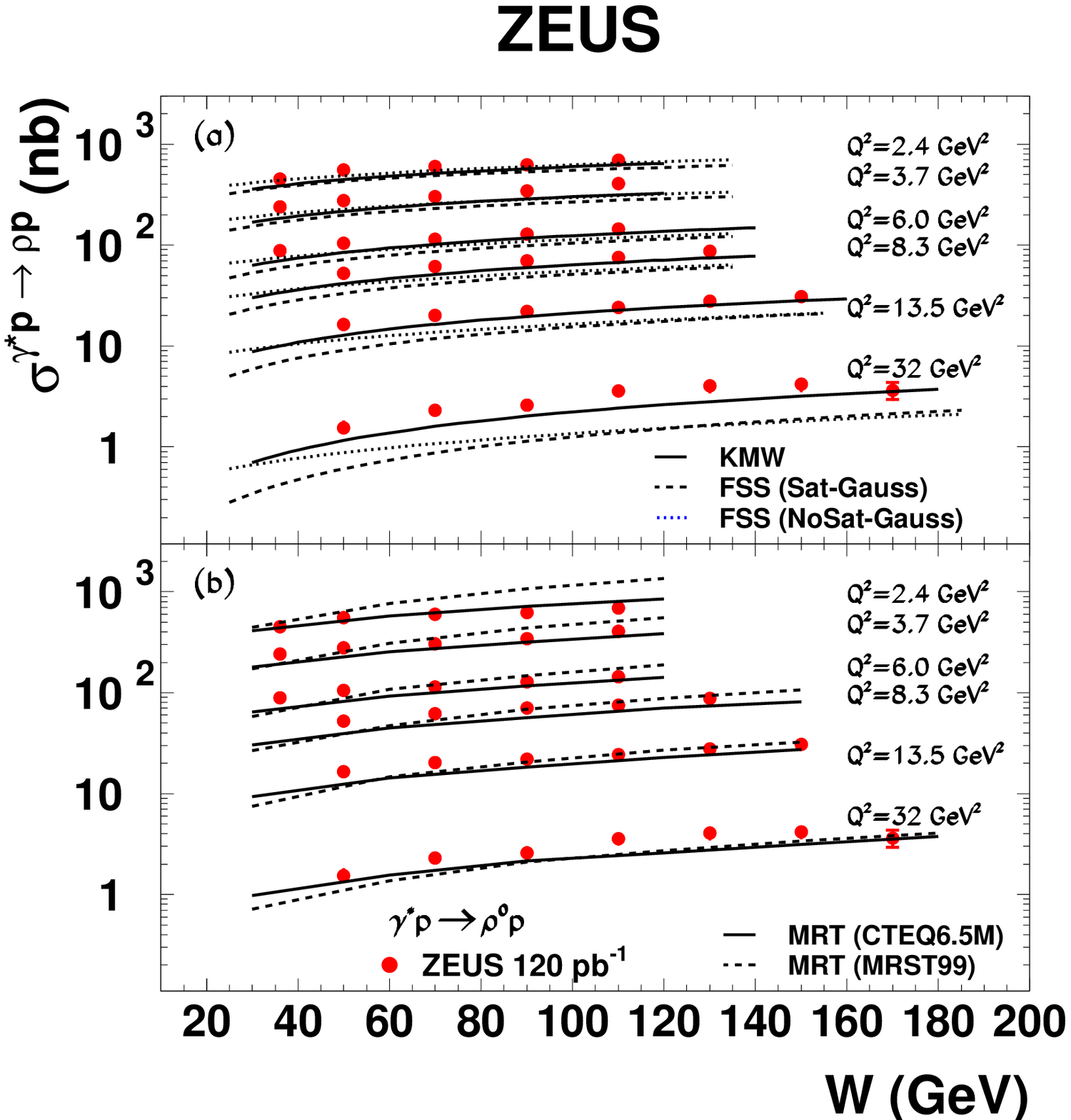}}
\vspace{-0.5cm}
\caption{\it 
The $W$ dependence of the $\gamma^* p \to \rho^0 p$ cross section for
different values of $Q^2$, as indicated in the figure.  The same data
are plotted in (a) and (b), compared to different models, as described
in the text. }

\label{fig:w-mod}
\end{minipage}
\end{figure}

The different predictions discussed above are compared to the $Q^2$
dependence of the cross section in Fig.~\ref{fig:q2-mod}.  None
of the models gives a good description of the data over the full
kinematic range of the measurement. The FSS model with the
three-dimensional Gaussian $\rho^0$ wave-function describes the
low-$Q^2$ data very well, while the KMW and DF models describe
the $Q^2>1$ GeV$^2$ region well.

The various predictions are also compared with the $W$ dependence of
the cross section, for different $Q^2$ values, in
Fig.~\ref{fig:w-mod}.  Here again, none of the models reproduces
the magnitude of the cross section measurements. The closest to the
data, in shape and magnitude, are the MRT model with the CTEQ6.5M
parametrisation of the gluon distribution in the proton and the KMW
model. 

\begin{figure}[h]
\begin{minipage}{0.5\columnwidth}
\hspace{-0.5cm}
\centerline{\includegraphics[width=\columnwidth]{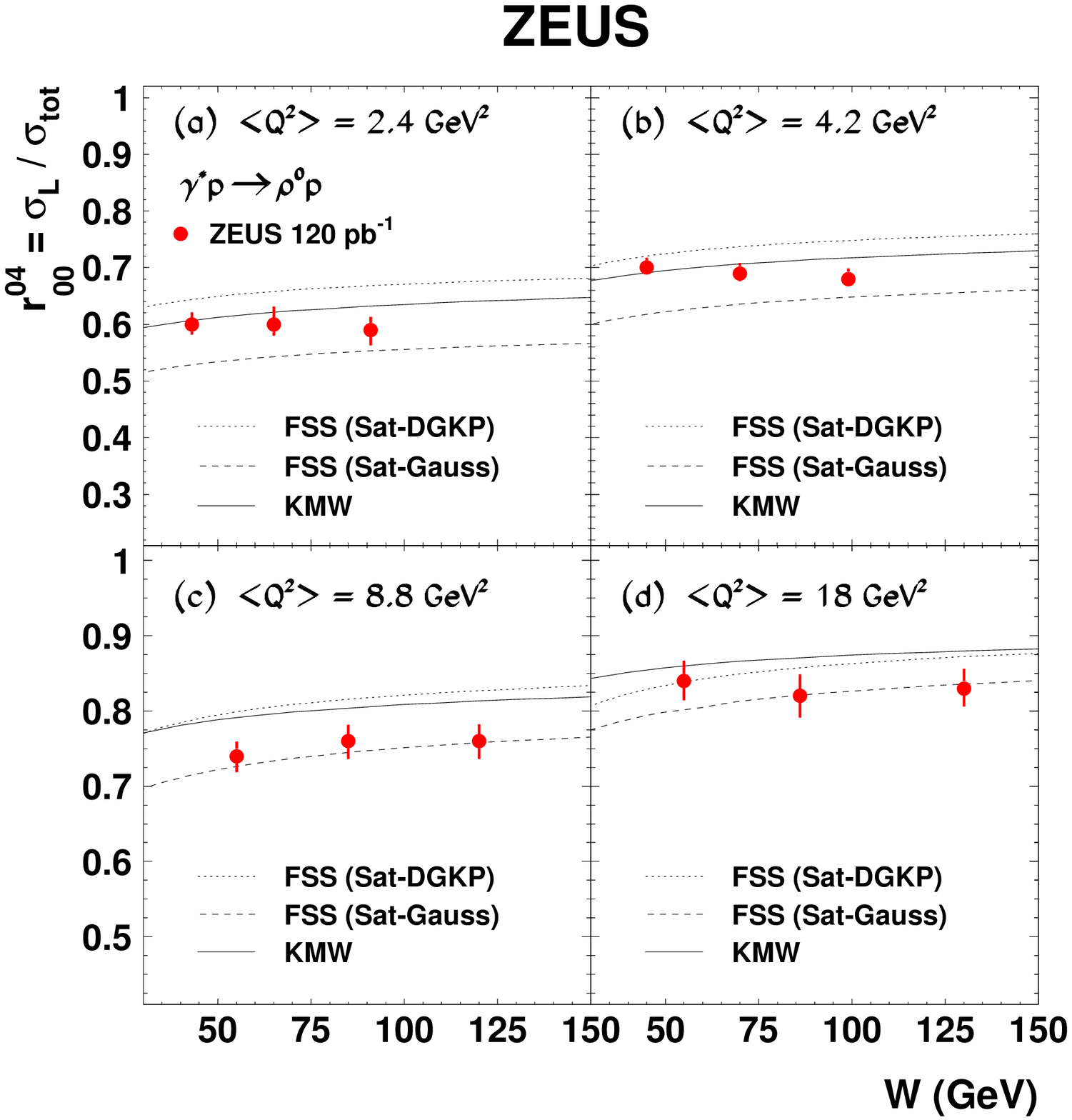}}
\vspace{-0.5cm}
\caption{\it
The ratio $r^{04}_{00}$ as a function of $W$ for different values of
$Q^2$ compared to the predictions of models as indicated in the figure
(see text).}
\label{fig:r-mod1}
\end{minipage}
\hspace{2mm}
\begin{minipage}[h]{0.5\columnwidth}
\hspace{-0.5cm}
\centerline{\includegraphics[width=\columnwidth]{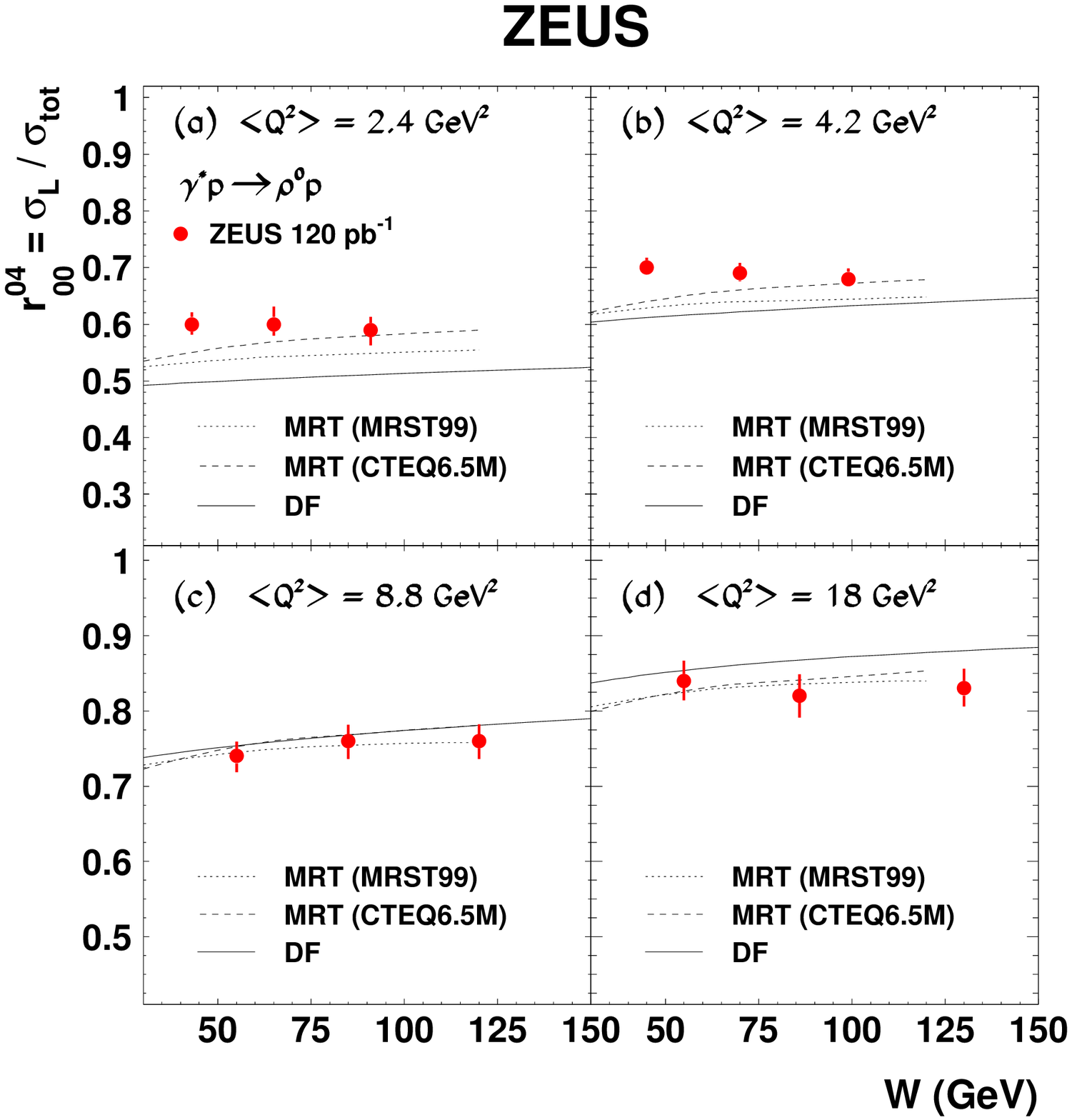}}
\vspace{-0.5cm}
\caption{\it
The ratio $r^{04}_{00}$ as a function of $W$ for different values of
$Q^2$ compared to the predictions of models as indicated in the figure
(see text).}
\label{fig:r-mod2}
\end{minipage}
\end{figure}

While all the models exhibit a mild dependence of $r^{04}_{00}$ on
$W$, consistent with the data as shown in Figs.~\ref{fig:r-mod1}
and~\ref{fig:r-mod2}, none of them reproduces correctly the magnitude
of $r^{04}_{00}$ in all the $Q^2$ bins.

In summary, none of the models considered above is able to describe
all the features of the data. The high precision of the measurements
can be used to refine models for exclusive $\rho^0$ electroproduction
and contribute to a better understanding of the $\rho^0$ wave function
and of the gluon density in the proton.


\begin{footnotesize}


\end{footnotesize}
\end{document}